\documentclass[twocolumn,twoside,slac_two]{revtex4}
\usepackage{graphicx}
\usepackage{fancyhdr}
\begin{document}

\title{Double-humped Super-luminous Supernovae}

%

\author{D. Leahy, R. Ouyed}
\affiliation{University of Calgary, 2500 University Dr NW, Calgary, AB, T2N1N4, Canada}
%

\begin{abstract}
Super-luminous supernova (SLSN) are supernovae showing extreme properties in their light-curves: high peak luminosities (more than 10 times brighter than bright SN Ia), and long durations. 
Several mechanisms have been proposed for SLSN, such as pair instability SN of a massive progenitor, interaction of the ejecta with a massive circumstellar shell, and the dual-shock
quark nova (dsQN) model. The dual-shock quark nova model is unique
in that it predicts a normal SN event will be seen $\sim$10 days
prior to the main SLSN event. The dsQN model is described here
and shown that it is consistent with the light curve of the one currently known double-humped SLSN, 2006oz.  

\end{abstract}

\maketitle

\thispagestyle{fancy}


\section{SUPER-LUMINOUS SUPERNOVAE}

Super-luminous supernovae (SLSNe) have been discovered in significant numbers over the past decade. 
They have high peak brightness, with peak absolute magnitudes of -21 or brighter, and usually have long durations (up to 1 year) compared to normal type I or type II SN. 
Mechanisms proposed to produce such high luminosities for so long 
include: pair instability SN (which requires a massive progenitor, 
$>100 M_{\odot}$); the interaction of fast moving SN ejecta ($\sim 
10M_{\odot}$) with a massive circumstellar shell (also with $\sim 
10M_{\odot}$); and the dual-shock quark nova (dsQN) model. 
The quark nova (QN) was proposed as an explanation for SN 2006gy 
and other SLSNe including SN 2005ap ~\cite{Leahy1}.
In  ~\cite{ouyed09a}, 
we emphasize that the lightcurve of the preceding SN gives 
a double-humped lightcurve. 

In the dsQN model (see section 2 below for details), a normal core-collapse SN explodes to produce a 
high-mass neutron star. The neutron star converts to a quark star, 
with a delay of several days, in a violent explosion called a quark 
nova(QN). The shock produced by the QN then reheats the SN ejecta, 
which can radiate at high levels for extended periods of time- 
because the SN ejecta has already expanded so that adiabatic 
expansion losses are much slower than for the case of a normal SN.
It is seen that the dual-shock quark nova model has unique
signature: it predicts that a normal core-collapse SN event will be 
seen several days prior to the main SLSN event. 

There is currently one known double-humped SLSN, namely 2006oz
~\cite{leloudas}.  
Supernova (SN) 2006oz  is a 
newly-recognized member of the class of H-poor, superluminous
supernovae ~\cite{Quimby}. 
The bolometric light curve shows a precursor event with a
duration between 6-10 days in the rest-frame, 
followed by a dip, after which the luminosity begins to
rise. The subsequent rise has previously been fit using three different models: 
(i) input from radioactive decay; (ii) a magnetar spin-down model; (iii) a circum-stellar medium
(CSM) interaction. The Nickel decay model has problems
because it requires an unreasonably large amount (10.8$M_{\odot}$) of $^{56}$Ni with a total ejecta mass of 14.4$M_{\odot}$. 
Another problem is that the SN was not detected 9 months later, 
which is inconsistent with the
standard decay curve for $^{60}$Co. The magnetar and CSM
lightcurve models were shown in Figure 7 of  ~\cite{leloudas}. 
None of these three models accounts for the precursor event.
Yet the dsQN model for SLSN, predicts the existence of a precursor
SN  ~\cite{ouyed09a} very similar to the observed precursor of SN2006oz.

To show the precursor of SN2006oz is plausibly a normal SN, we estimate 
its energy be $\sim 10^{49}\ {\rm erg}\times t_{\rm pre, 10}$ where
   $t_{\rm pre,10}$ is the duration of the precursor in units of 10 days (limited by the observations from about 7 days
   to 12 days).  This energy is typical of  brighter Type-II SNe (e.g. ~\cite{Young}).  
  
In this paper we describe the dual-shock QN (dsQN) model, including the precursor SN. Then we show that
the main peak and the precursor of SN2006oz are self-consistently
fit by the dsQN model, and conclude with remarks including future work to be done.
 
\section{THE DUAL-SHOCK QUARK NOVA (dsQN) MODEL}


A quark nova (QN) was proposed as an alternative
explanation for SN 2006gy ~\cite{Leahy1},
~\cite{ouyed09a}. A QN is expected to occur when
the core density of a neutron star reaches the quark deconfinement
density and triggers a violent ~\cite{ouyed02}
conversion to the more stable strange quark matter
~\cite{Itoh}, ~\cite{Bodmer}, ~\cite{Witten}.
During the spin-down evolution
of the neutron star, accompanied by increasing central density, a detonative ~\cite{Niebergal}, ~\cite{Ouyed11}
phase transition to up-down-strange
triplets would result in ejection of the outer heavy element-rich and neutron-rich layers of the neutron
star at ultra-relativistic velocities ~\cite{Keranen}, ~\cite{OL09}
(in ~\cite{OL09}, see the first panel of Fig. 2).
 Follow-up studies of neutrino and photon emission
processes during the QN  ~\cite{Vogt}, ~\cite{Ouyed09b}
have shown that these outermost layers
(of $\sim 10^{-4}$-$10^{-3} M_{\odot}$ in mass) can be ejected with up to $10^{53}$  erg in
kinetic energy. Nucleo-synthesis simulations of the evolution
of the neutron-rich QN ejecta were found to
produce primarily heavy elements with mass number, A
$>$ 130  ~\cite{Jaikumar}.

If the time delay($t_{\rm delay}$) 
between SN and QN explosions is too long the SN ejecta will
have dissipated such that the QN essentially erupts in
isolation. However, when $t_{\rm delay}$ is on the order of days to
weeks a violent collision occurs reheating the extended SN ejecta 
~\cite{Leahy1}, ~\cite{ouyed09a}.
The brilliant radiance of the re-shocked SN ejecta fades as
the photosphere recedes, eventually revealing a mixture
of the inner SN ejecta and the QN ejecta material.

\begin{figure*}[t]
\centering
\includegraphics[width=135mm]{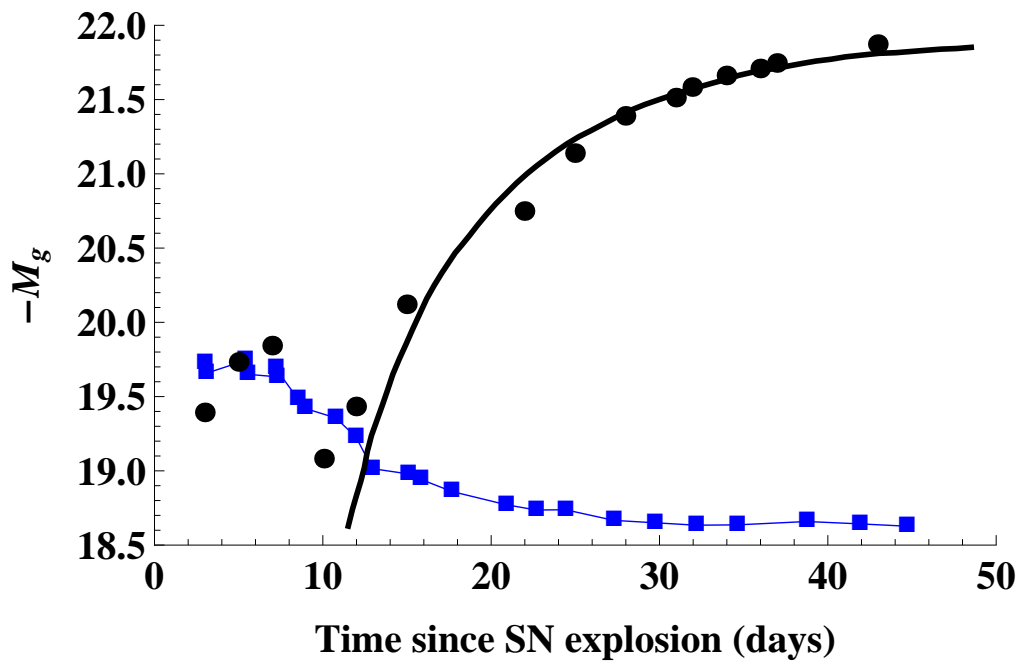}
\caption{(reproduced from ~\cite{OL12}) SN2006oz r-band lightcurve (solid circles; upper limits
shown as triangles). The dsQN model is calculated for
$M_{\rm ejecta}=20M_{\odot}$ and ($t_{\rm delay}$) = 6.5 days.} \label{fig1}
\end{figure*}



\section{A PLAUSIBLE CANDIDATE DOUBLE-HUMPED dsQN EVENT: SN2006oz}

SN2006oz is the first known double-humped SLSN event ~\cite{leloudas}.
~\cite{OL12} study this event is some detail. Here we use SN2006oz
as an example of double-humped SNSNe, and argue that, in the dsQN model, we expect to see other similar examples in future.

Figure 1 (from ~\cite{OL12})) shows the observed SN2006oz light curve using the data from ~\cite{leloudas}.
The g-band data is used, which has the best time coverage and smallest errors for most times.
Time is in days at the source using the known redshift ($z=0.376$). 
Apparent g-band magnitudes were converted to absolute g-band magnitudes using the 
corresponding luminosity distance for the standard model 
~\cite{Wright}. 
The suggested extinction correction (B-V) from 
 ~\cite{leloudas}  was included, even though it was small. 
  The dsQN model is also shown in Fig. 1.  For the
  SN lightcurve (the first hump), we use an observed light curve:
 the light curve of SN1999em from ~\cite{Bersten}
  which has good time coverage in the first 50 days. ~\cite{Bersten11}
  fitted hydrodynamic models to SN1999em and derived a progenitor mass of $19M_{\odot}$ which is similar
  in mass to the SN progenitor we used in our QN model.
Other parameters for the SN1999em model were progenitor radius of $800R_{\odot}$, explosion energy of $1.25\times 10^{51}$ erg and 
  $^{56}$Ni mass of 0.056$M_{\odot}$. 
 We scaled the bolometric magnitude by +2 to  
   represent a more energetic SN, which is reasonable  
    since the range in brightness of Type II SNe
  varies  considerably with many models giving brighter SN than 1993em (e.g.~\cite{Young}).

In the QN model the progenitor initial mass is in the range of 20-40$M_{\odot}$ (see ~\cite{Leahy1},~\cite{Ouyed09b},~\cite{Ouyed10})
to create a massive neutron star with core density near the instability to convert to quark matter ~\cite{Niebergal}.
This motivates our choice of SN  ejected mass of 20$M_{\odot}$. Best fits from our previous studies
of SLSNe yielded time delays of $\sim 10$ days which motivates the time delays that
we explored. For SN2006oz the shown fit (see Figure 1) uses $t_{\rm delay}=6.5$ days,
$v_{\rm QN}= 5000$ km s$^{-1}$ and a preceding SN ejecta with an average velocity of $v_{\rm SN}\simeq 1900$ km s$^{-1}$.
The combined light from the SN and from the QN-reheated SN ejecta give a reasonable
 fit to the observations with a self-consistent model.

\section{COMMENTS}

~\cite{leloudas} notes the intriguing
possibility of an intrinsic precursor event in SN
2005ap-like objects. In the dsQN model, there must be
a normal SN (−20 $<$ Mbol $<$ −15) preceding the
SLSN (if the delay is long enough, $>\sim$ 10 days) which
should be detectable for nearby SN 2005ap-like explosions.
For short delays, the normal SN lightcurve would overlap with the brighter dsQN lightcurve and not give a distinct hump.

SN 2005ap-like events are rare: 
they occur at a rate of less than one in $10^4$
core-collapse SNe  ~\cite{Quimby}. 
dsQNe are also expected to be rare events: the QNe rate is estimated
to be $\sim$ one in 1000 core-collapse events with one tenth of
them having time delays in the appropriate range
to produce dsQNe (tdelay $\sim$ 5-30 days (~\cite{Staff},
~\cite{Jaikumar}, ~\cite{Leahy1}, ~\cite{Leahy09}, ~\cite{ouyed09a}).
These two order
of magnitude estimates of dsQN events and 
SN 2005ap-like events are consistent with eachother.

We note that the dsQN model applies to both H-rich and H-poor
SLSNe, but these occur at similar rates. For both cases, the QN shock reheats the SN envelope to high temperature,
so H-poor/H-rich progenitors would give H poor/
H-rich spectra. Because of mass-loss dependency on metallicity, 
we expect H poor SLSNe to occur in higher-metallicity environments.
Low-metallicity progenitors would more likely be H-rich and 
should in principle have more massive envelopes. 

We expect a number of SLSNe to have a double-humped character, as 
predicted by the dsQN model. The first hump is much fainter and has 
the brightness of a normal core-collapse SN, but should be 
observable in relatively nearby SLSNe. It is an for the dsQN model 
to find additional double-humped SLSNe beyond SN2006oz. 
We can model the precursor light-curves to learn about the progenitors of SLSNe and of dsQN. 
The precursor SN is also a clear and unique signature of the dsQN model. 
Other properties predicted by the dsQN model include: the presence of heavy elements ~\cite{Jaikumar}, and of spallation nuclei produced in the collision between the fast-moving QN ejecta and the inner parts of the SN ejecta ~\cite{Ouyed11}.

\bigskip 
\begin{acknowledgments}
This research is supported the National Science and Engineering Research Council of Canada (NSERC).
\end{acknowledgments}

\bigskip 

\end{document}